\documentclass[prl,aps,superscriptaddress,twocolumn,notitlepage,showpacs]{revtex4-1}
\usepackage{latexsym}
\usepackage{amsmath}
\usepackage{amssymb}
\usepackage{graphicx}
\usepackage{caption}
\usepackage{subfigure}
\usepackage{float}
\usepackage{mathrsfs}
\usepackage{color}
\usepackage{txfonts}
\usepackage[justification=centering,
            format=plain]{caption}

\renewcommand{\raggedright}{\leftskip=0pt \rightskip=0pt plus 0cm}

\begin{document}

\title{Quantum fluctuations in Fr\"ohlich condensate of molecular vibrations driven far from equilibrium}

\author{Zhedong Zhang}
\email{zhedong.zhang@tamu.edu}
\affiliation{Institute for Quantum Science and Engineering, Texas A$\&$M University, College Station, TX 77843, USA}

\author{Girish S. Agarwal}
\email{girish.agarwal@tamu.edu}
\affiliation{Institute for Quantum Science and Engineering, Texas A$\&$M University, College Station, TX 77843, USA}
\affiliation{Department of Biological and Agricultural Engineering, Department of Physics and Astronomy, Texas A$\&$M University, College Station, TX 77843, USA}

\author{Marlan O. Scully}
\email{scully@tamu.edu}
\affiliation{Institute for Quantum Science and Engineering, Texas A$\&$M University, College Station, TX 77843, USA}
\affiliation{Quantum Optics Laboratory, Baylor Research and Innovation Collaborative, Waco, TX 76704, USA}
\affiliation{Department of Mechanical and Aerospace Engineering, Princeton University, Princeton, NJ 08544, USA}

\date{\today}

\begin{abstract}
Fr\"ohlich discovered the remarkable condensation of polar vibrations into the lowest frequency mode when the system is pumped externally. For a full understanding of the Fr\"ohlich condensate one needs to go beyond the mean field level to describe critical behavior as well as quantum fluctuations. 
The energy redistribution among vibrational modes with nonlinearity included is shown to be essential for realizing the condensate and the phonon-number distribution, revealing the transition from quasi-thermal to super-Poissonian statistics with the pump. We further study the spectroscopic properties of the Fr\"ohlich condensate, which are especially revealed by the narrow linewidth. This gives the long-lived coherence and the collective motion of the condensate. Finally we show that the proteins such as Bovine Serum Albumin (BSA) and lysozyme are most likely the candidates for observing such collective modes in THz regime by means of Raman or infrared (IR) spectroscopy. 
\end{abstract}

\maketitle

{\it Introduction.--} The collective properties in both physical and biological systems attracted much attention during past decades and are of great importance for understanding many peculiar phenomenons, such as nonequilibrium phase transition of polaritons \cite{Kasprzak_Nature2006,Amo_NatPhys2009,Donner_PRL2018}, superefficient energy transfer in photosynthesis \cite{Engel_Nat2007,Zhang_SR2016,Ishizaki_ARCM2012,Zhang_JPCB2015} and cognitive function of some molecular machinery at work in living cells \cite{Hameroff_1998,Hagan_PRE2002}. In order to understand such out-of-equilibrium collective feature, Fr\"ohlich suggested a condensation of energy at the lowest mode of polar vibrations once the external energy supply exceeds a threshold \cite{Frohlich_1968} and this idea was further followed by others for detailed investigation over three decades \cite{Tuszynski_PRA1984,Tuszynski_PRE2001,Pokorny_JTB1982,Mesquita_IJQC2005}. Provided the sufficient energy pump, this large accumulation of phonons considerably builds up at the lowest mode. This is a reminiscence of Bose-Einstein condensate (BEC) \cite{Ketterle_PRL1995,Jin_PRL1996,Pethick_book2001}. A remarkable progress was recently made by the numerical simulations based on Wu-Austin Hamiltonian \cite{Wu_JBP1981}, which specified the parameter regime for weak, mediate and strong condensates in various proteins \cite{Reimers_PNAS2009}.

To achieve such out-of-equilibrium condensate, the energy redistribution essentially plays an important role of introducing the nonlinearity \cite{Frohlich_1968,Reimers_PNAS2009}. The mechanism of the emergence of Fr\"ohlich condensate reveals the analogy to the laser operation \cite{Scully_PRL1999,Scully_PRA1970}, which offers new insight to understand the cooperative phenomena in chemical and biological systems \cite{Dorfman_PNAS2013}. The coherent nature of laser is manifested by the narrow linewidth. Thereby, under proper conditions the Fr\"ohlich condensate would show the coherent feature which is one of the main tasks of this work. It may be noted that the redistribution of energy by the pumping mechanism could allow for the control of reaction kinetics \cite{comments}.

The combination of recent advance on X-ray crystallography \cite{Miao_Nat1999} and THz radiation offer effective tools for visualizing the structural change associated with low-frequency collective vibrations in the materials, i.e., lysozyme protein crystals \cite{Lundholm_SD2015,Turton_NatCommun2013}. The structural change has been observed to sustain for micro- to milli-seconds, which is several orders of the magnitude longer than the one induced by the redistribution of THz-vibrations towards thermal distribution \cite{Lundholm_SD2015}. It seems plausible to attribute this phenomena to Fr\"ohlich condensate that causes non-thermal distribution. The most recent experiment demonstrated in BSA protein the remarkable absorption feature around 0.314THz, when driving the system by optical pumping \cite{Nardecchia_2017}. Understanding these interesting and intriguing experiments will lead us to the deep thinking about Fr\"ohlich's mechanism, since the cooperativity has been shown to exist in some non-physical systems \cite{Haken_RMP1975,Zhang_JPCL2017}. On the other hand, this out-of-equilibrium cooperativity manifests its analogy even at classical level, such that the long-survived limit cycle oscillation in gene network when increasing the binding of active protein to the gene \cite{Wolynes_PNAS2013,Wang_PNAS2014}. Thus it is important to obtain a detailed understanding of the coherence of Fr\"ohlich condensate, which is still obscured and short of the direct evidence, especially in THz regime. 
Clearly a full understanding can solely come from a quantum theory which would give not only the mean but also quantum fluctuations as well as line-width information, crucial in spectroscopic measurement.

In this Letter we develop a full quantum statistical theory for Fr\"ohlich condensate. We analytically find that the phonon-number distribution evolves with external energy pump, from quasi-thermal to super-Poissonian statistics (sub-Poissonian statistics is shown in Supplementary Material). 
Besides, the long-lived coherence of the condensate is observed, i.e., the dramatic increase of the lifetime by $\sim 15$ times for BSA under room temperature ($\sim 50$ times under low temperature). This subsequently characterizes the narrow linewidth in the spectroscopic signal. We suggest some possible candidates for observing the Fr\"ohlich condensate, such as lysozyme and BSA proteins rather than the longitudinal vibrations of microtubules. Moreover the phonon statistics paves the road for evaluating the quantum fluctuations in the condensate particles. Note that the experimental studies of the quantum fluctuations in condensates such as exciton polariton condensate have been reported recently \cite{Klaas_PRL2018}.

\begin{figure}
 \captionsetup{justification=raggedright,singlelinecheck=false}
 \centering
   \includegraphics[scale=0.22]{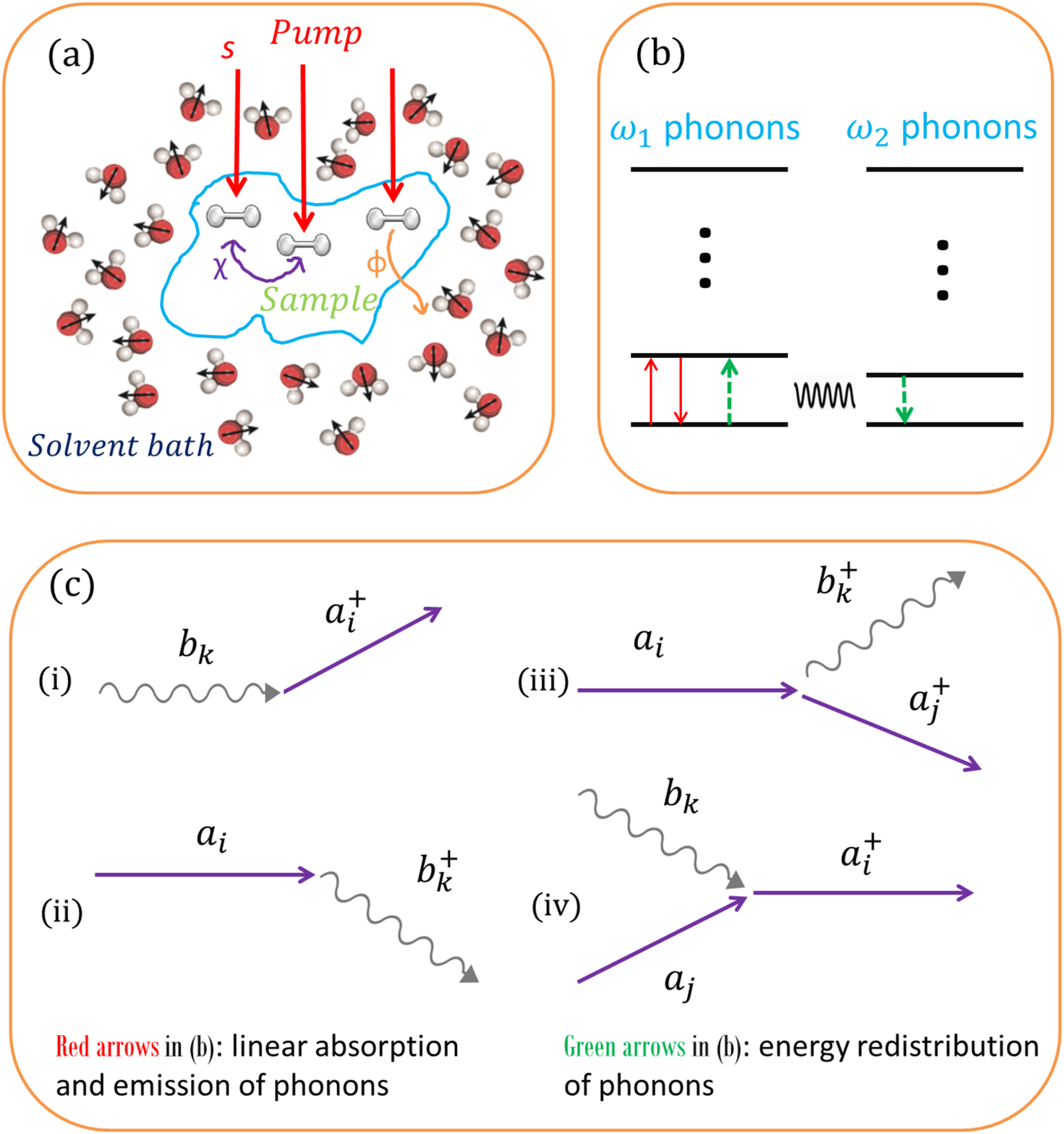}
\caption{(a) Schematic of molecular vibrations driven into far-from-equilibrium regime; (b) Vibrational modes absorb energy from external source at rate of $s$ and surrounding medium acting as thermal bath causes the dissipation at rate of $\phi$, corresponding to the 1st term in Eq.(\ref{Vp}). Bath is also responsible for the energy redistribution (nonlinear) at rate of $\chi$, corresponding to the 2nd term in Eq.(\ref{Vp}); (c) Feynman diagrams for those 1st and 2nd order processes.} 
\label{sch}
\end{figure}

{\it Model and equation of motion.--} For the low-frequency vibrations in molecules, i.e., intramolecular vibrations of proteins in THz regime and low-energy phonon modes in DNA, the surrounding medium, such as solvent/water, acts as a thermal environment, resulting in the energy dissipation and redistribution. The latter shows the nonlinearity which plays an important role in biological functions involving the low-loss energy transport through protein as attributed to the underdamped low-frequency phonon modes in DNA \cite{Dabydov_JTB1973}. Besides, these vibrational modes are excited by a continuous energy supply, as shown in Fig.\ref{sch}. We may model the system as a group of harmonic oscillators with the frequency spectrum $\omega_j,\ (j=0,1,2,\cdots,D)$ where $D\gg 1$ denotes the total number of vibrational modes. The effect of external pump and environment is governed by the coupling term $V(t)=V_{\text{p}}(t)+\left(V_{\text{env}}(t)+\text{h.c.}\right)$ where $V_{\text{p}}(t) = \sum_{s=0}^D\left(F_s^*(t)a_s e^{-i\omega_s t} + \text{h.c.}\right)$ and
\begin{equation}
\begin{split}
V_{\text{env}}(t) = \sum_k\left(\sum_{s=0}^D \frac{f_{s,k}}{\sqrt{2}}a_s^{\dagger}b_k e^{i\Delta_s^k t} + \sum_{s>l\ge 0}^D \frac{G_{sl,k}}{\sqrt{2}}a_s^{\dagger}a_l b_k e^{i\Delta_{sl}^k t}\right)
\end{split}
\label{Vp}
\end{equation}
The pumping field is characterized by a broad spectrum: $\langle F_s(t)F_l^*(t')\rangle=\frac{r_s}{2}\delta_{sl}\delta(t-t')$ with a pumping rate of $r_s$ and $\Delta_s^k=\omega_s-v_k,\ \Delta_{sl}^k=\omega_s-\omega_l-v_k$. $a_s$ and $b_k$ stand for the bosonic annihilation operators for molecular vibrations and bath modes, respectively. The first term in Eq.(\ref{Vp}) describes the energy exchange between the vibrational modes and environment, giving the dissipation (one-phonon process). The second term in Eq.(\ref{Vp}) quantifies the two-phonon process, causing energy redistribution between the vibrational modes \cite{Wu_JBP1981}. Defining the density matrix $\rho_{n_0,m_0} = \sum_{\{n_l\}}\langle n_0;\{n_l\}|\rho|m_0;\{n_l\}\rangle$ for the mode $\omega_0$, the equation of motion may be derived, by adopting the tools from quantum optics \cite{Agarwal_book2013,Scully_book1997}. The calculation consists of two steps: (1) derivation of the master equation for the density matrix of phonons by eliminating the solvent degrees of freedoms; (2) use of the phonon master equation to obtain a reduced description of the lowest phonon frequency mode. 
All the details are given in Supplementary Material (SM). The full master equation results in the equation which the total phonon number $N=\sum_{s=0}^D \langle a_s^{\dagger}a_s\rangle$ obeys: $\dot{N}=(D+1)(r+\phi\bar{n})-\phi N$ showing that the total phonon number is {\it solely} dictated by the external pumping and dissipation. $r$ and $\phi=2\pi f_{\omega}^2{\cal D}(\omega)$ refer to the rates of energy pumping and dissipation, respectively. $\bar{n}=[\text{exp}(\hbar\omega_0/k_B T)-1]^{-1}$ is the Planck factor. Thus $N$ does not depend on the processes (iii) and (iv) in Fig.\ref{sch}(c). We can in fact replace $N$ by its stationary value as the timescale of interest $t>\phi^{-1}$. Thereby $N$ can be partitioned into $N=N_{\text{r}}+N_{\text{th}}$ where $N_{\text{r}}=(D+1)r/\phi$ and $N_{\text{th}}=(D+1)\bar{n}$. This manifests the contributions to the total excitation from external pumping $r$ and thermal distribution $\bar{n}$. Such observation enables us to derive a simple equation for the population at the lowest vibrational mode $\omega_0$, assuming $\langle n_0 N\rangle\simeq\langle n_0\rangle N$

\begin{figure}
 \captionsetup{justification=raggedright,singlelinecheck=false}
 \centering
   \includegraphics[scale=0.22]{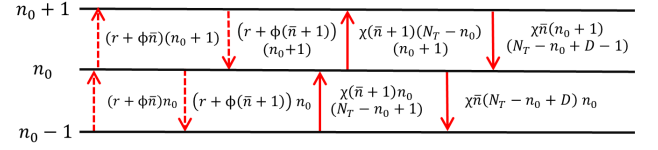}
\caption{Transition diagram in accordance to Eq.(\ref{qmecon}), showing the flow of probability in and out of the $|n_0\rangle$ state from and to the neighboring $|n_0+1\rangle$ and $|n_0-1\rangle$ states. $(r+\phi\bar{n})(n_0+1)$ and $(r+\phi(\bar{n}+1))(n_0+1)$ represent the respective phonon emission and absorption, due to the one-phonon processes (pumping and dissipation); $\chi(\bar{n}+1)(N_T-n_0)(n_0+1)=\phi\bar{n}(n_0+1)\left(\chi(\bar{n}+1)/\phi\bar{n}\right)(N_T-n_0)$ corresponds to the process in which phonons are emitted owing to both dissipation and energy redistribution; $\chi\bar{n}(N_T-n_0+D-1)(n_0+1)=\phi(\bar{n}+1)(n_0+1)\left(\chi\bar{n}/\phi(\bar{n}+1)\right)(N_T-n_0+D-1)$ corresponds to the process in which phonons are absorbed owing to both dissipation and energy redistribution. Similar explanations exist for the other terms.}
\label{trans}
\end{figure}

\begin{figure}
 \captionsetup{justification=raggedright,singlelinecheck=false}
 \centering
   \includegraphics[scale=0.14]{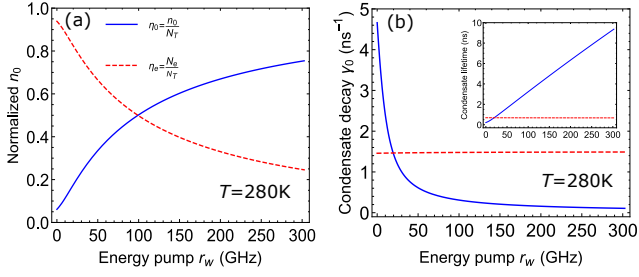}
\caption{(a) Phonon numbers at the lowest mode $\omega_0$ and excited modes $\omega_l\ (l\ge 1)$ normalized by total phonon number $N$ vary with external energy supply; (b) Damping rate $\gamma_0$ of condensate varies with external energy supply. Lifetime $1/\gamma_0$ is plotted in small panel. In (b), dashed red line is for the case with no Fr\"ohlich condensate when turning off the nonlinearity. Solvent temperatures $T=280$K, corresponding to $\bar{n}=16$. Parameters are taken from Ref.\cite{Nardecchia_2017} for BSA protein: $\omega_0=0.314\times 2\pi$THz, $\phi=6$GHz, $\chi=0.07$GHz and $D=200$.}
\label{n0fig}
\end{figure}

\begin{equation}
\begin{split}
& \dot{\rho}_{n_0,n_0} = -\big(r+\phi\bar{n}+\chi{\cal N}_{n_0}\big)(n_0+1)\rho_{n_0,n_0}\\[0.15cm]
& \qquad\quad + \big(r+\phi\bar{n}+\chi{\cal N}_{n_0-1}\big)n_0\rho_{n_0-1,n_0-1}\\[0.15cm]
& \qquad\quad - \big(r+\phi(\bar{n}+1)+\chi{\cal M}_{n_0}\big)n_0\rho_{n_0,n_0}\\[0.15cm]
& \qquad\quad + \big(r+\phi(\bar{n}+1)+\chi{\cal M}_{n_0+1}\big)(n_0+1)\rho_{n_0+1,n_0+1}
\end{split}
\label{qmecon}
\end{equation}
from the reduced master equation, where ${\cal N}_{n_0}=\sum_{j=1}^D(\bar{n}_{\omega_{j0}}+1)\langle n_j\rangle_{n_0},\ {\cal M}_{n_0}=\sum_{j=1}^D\bar{n}_{\omega_{j0}}\langle n_j+1\rangle_{n_0}$. There are varying degrees of rigor to evaluate ${\cal N}_{n_0}$ and ${\cal M}_{n_0}$. We will choose the one assuming $\bar{n}_{\omega_{j0}}\simeq\bar{n}$ which leads to ${\cal N}_{n_0}\simeq(\bar{n}+1)(N-n_0),{\cal M}_{n_0}\simeq\bar{n}(N-n_0+D)$. Eq.(\ref{qmecon}) has a nice interpretation in terms of the probability flows as depicted in Fig.\ref{trans}.

{\it Out-of-equilibrium condensation of phonons.--} To illustrate the condensation of phonons and its critical behaviors, we essentially obtain the rate equations for the phonon number $\langle n_0\rangle$ at the lowest mode
\begin{equation}
\begin{split}
\langle \dot{n}_0\rangle = \left(\chi N_{\text{r}}-\phi-\chi\right)\langle n_0\rangle - \chi\langle n_0^2\rangle + \left[r+\phi\bar{n}+\chi(\bar{n}+1)N\right]
\end{split}
\label{n0}
\end{equation}
Each term has a physical meaning. The mode $\omega_0$ experiences gain due to the pumping of all the modes and loses energy via the term $(\phi+\chi)\langle n_0\rangle$ as well as the nonlinear term $\chi\langle n_0^2\rangle$. The last bracket in Eq.(\ref{n0}) lead to the residue number of phonons in mode $\omega_0$ even below the pumping threshold
\begin{equation}
\begin{split}
r_{\text{c}} = \frac{\phi}{D+1}\left(1+\frac{\phi}{\chi}\right)
\end{split}
\label{rc}
\end{equation}
as indicated by $\chi N_{\text{r}}-(\phi+\chi)>0$ in Eq.(\ref{n0}). The equation for $\langle n_0\rangle$ has a structure which is a reminiscence of the photon number for a single mode laser.

Further simplification neglects the fluctuation of $n_0$, namely, $\langle n_0^2\rangle\simeq \langle n_0\rangle^2$, which gives the steady-state phonon number at mode $\omega_0$ shown in Fig.\ref{n0fig}(a) where the parameters are taken from recent experiment \cite{Nardecchia_2017}. Moreover, the expansion $\langle n_0\rangle$ in terms of $r-r_{\text{c}}$ leads to the scaling $\langle n_0\rangle\simeq F_0+F_1(r-r_{\text{c}})$ when approaching the critical point $r=r_{\text{c}}$. This gives the critical exponent $\beta=1$.

\begin{figure}
 \captionsetup{justification=raggedright,singlelinecheck=false}
 \centering
   \includegraphics[scale=0.195]{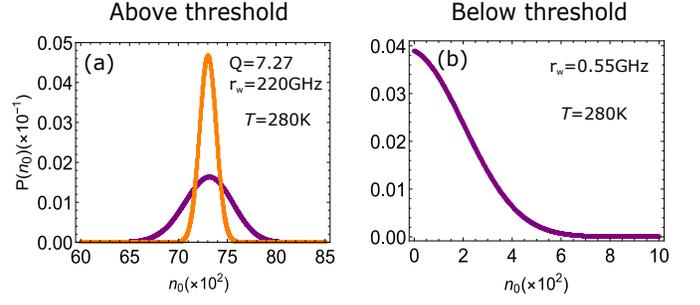}
\caption{Phonon distribution of the lowest vibrational mode where the solvent temperature is $T=280$K corresponding to $\bar{n}=16$. Energy pump (a) $r=220$GHz, (b) $r=0.55$GHz; Orange dots correspond to Poissonian distribution. Other parameters are the same as that in Fig.\ref{n0fig}.}
\label{phononswt300}
\end{figure}

{\it Coherence of condensate and linewidth.--} As a collective mode, the in-phase motion should be maintained in the condensate. This is featured by the long-lived or long-range coherence, which may be deeply connected to the so-called off-diagonal long-range order (ODLRO) in superconductivity and superfluidity \cite{Yang_RMP1962}. The coherence characteristics of our condensate are given by the dynamical evolution associated with the off-diagonal elements of density matrix $\rho_{n_0,n_0+1}$. To probe such coherence properties, we can use infrared or Raman spectroscopies. From the reduced master equation, the coherence dynamics obeys
\begin{equation}
\begin{split}
& \dot{\rho}_{n_0,n_0+1} = (i\omega_0-\gamma_{n_0})\rho_{n_0,n_0+1} + c_{n_0-1}\rho_{n_0-1,n_0}\\[0.1cm]
& \qquad\ \ - (c_{n_0}+d_{n_0})\rho_{n_0,n_0+1} + d_{n_0+1}\rho_{n_0+1,n_0+2}
\end{split}
\label{rhocoh}
\end{equation}
where
\begin{equation}
\begin{split}
\gamma_{n_0} = & \frac{1}{4}\Bigg[\frac{r+\phi(\bar{n}+1)+\chi\bar{n}(N-n_0+D)}{\sqrt{n_0(n_0+1)}+n_0+\frac{1}{2}}\\[0.15cm]
&\qquad + \frac{r+\phi\bar{n}+\chi(\bar{n}+1)(N-n_0)}{\sqrt{(n_0+1)(n_0+2)}+n_0+\frac{3}{2}}\Bigg]
\end{split}
\label{gamma0}
\end{equation}
and $c_{n_0} = \sqrt{(n_0+1)(n_0+2)}\ [r+\phi\bar{n}+\chi(\bar{n}+1)(N-n_0)]$, $d_{n_0} = \sqrt{n_0(n_0+1)}\ [r+\phi(\bar{n}+1)+\chi\bar{n}(N-n_0+D)]$. To solve Eq.(\ref{rhocoh}), we have the ansatz: $\rho_{n_0,n_0+1}(t) = \text{exp}[i\omega_0 t-D_{n_0}(t)]\rho_{0,1}(0)\prod_{m=1}^{n_0}c_{m-1}/d_m$ by imposing the detailed balance into the initial condition $c_{n_0}\rho_{n_0,n_0+1}(0)=d_{n_0+1}\rho_{n_0+1,n_0+2}(0)$ \cite{Scully_book1997}. The slow variation of $\gamma_{n_0}$ with respect to $t$ gives $|D_{n_0-1}-D_{n_0}|\ll 1$ yielding $D_{n_0}(t)\simeq \gamma_{n_0}t$. We can thereby safely replace $n_0$ in $\gamma_{n_0}$ by $\langle n_0\rangle$. Hence the vibrational coherence of the lowest mode is approximated to $\rho_{n_0,n_0+1}(t) \propto e^{(i\omega_0-\gamma_0) t}$ where the lifetime is given by
\begin{equation}
\begin{split}
\gamma_0\simeq \frac{r+\phi\left(\bar{n}+\frac{1}{2}\right)}{4\langle n_0\rangle}
\end{split}
\label{decay}
\end{equation}
for the energy supply much above threshold. Eq.(\ref{decay}) elucidates the considerable suppression of the damping rate by the large condensation of phonons. This is further supported by the numerical calculations displayed in Fig.\ref{n0fig}(b). It shows that the coherence lifetime when driving system into far-from-equilibrium regime ($r\simeq 200$GHz) is $\sim$10 times than the expected one induced by the redistribution of THz vibrations towards thermal distribution. 
This, in other words, will result in a remarkably sharp peak in the infrared and Raman spectroscopies. On the other hand, the suppression of damping rate by the large condensation of phonons implies the long-range coherence in such out-of-equilibrium condensate, due to $\ell_{\text{c}}\simeq v_s/\gamma_0$ where $v_s$ is the sound wave velocity ($\sim 1500$m/s in water). For BSA protein under room temperature, $\ell_{\text{c}}\simeq 1500\times (1/4.6)\times 10^{-9}=326$nm pumped at rate $r=0$ and $\ell_{\text{c}}\simeq 1500\times (1/0.3)\times 10^{-9}=5\mu$m pumped at rate $r=100$GHz. 

From the standard definition, we find the infrared fluorescence spectra governed by $\langle \mu^{(+)}(t)\mu^{(-)}(0)\rangle$ where $+(-)$ denotes the raising (lowering) part of the dipole operator
\begin{equation}
\begin{split}
S_{\text{FL}}(\omega) = \frac{32\pi^2 M\omega}{3\hbar\Omega}\sum_{j=0}^D\frac{|\mu_j|^2\gamma_j\langle n_j\rangle}{(\omega_j-\omega)^2+\gamma_j^2}
\end{split}
\label{xw}
\end{equation}
where $\gamma_j^{-1}$ stands for the lifetime of the $j$-th vibrational mode. $|\mu_j|$ refers to the magnitude of electric dipole moment of the $j$-th mode. $\Omega$ is the bulk volume and $M$ stands for the amount of molecules.
The $j=0$ term in Eq.(\ref{xw}) shows a high-intensity peak and a narrow linewidth associated with Fr\"ohlich condensate as supported by Fig.\ref{n0fig}(b). 

{\it Phonon statistics of Fr\"ohlich condensate.--} To gain more information about the out-of-equilibrium condensate of phonons obtained above, we proceed via the fluctuations of phonon number which is governed by the phonon distribution
\begin{equation}
\begin{split}
P(n_0) = P(0)\left(\frac{\alpha}{\beta}\right)^{n_0}\frac{\Gamma\left(\frac{\mathscr{X}}{\alpha}\right)\Gamma\left(\frac{\mathscr{Y}}{\beta}-n_0\right)}{\Gamma\left(\frac{\mathscr{X}}{\alpha}-n_0\right)\Gamma\left(\frac{\mathscr{Y}}{\beta}\right)}
\end{split}
\label{Pn0r}
\end{equation}
at steady state where $P(n_0)\equiv\rho_{n_0,n_0}^{ss}$. $\mathscr{X} = r + \phi\bar{n} + \chi(\bar{n}+1)(N+1),\ \mathscr{Y} = r + \phi(\bar{n}+1) + \chi\bar{n}(N+D)$ and $\alpha = \chi(\bar{n}+1),\ \beta = \chi\bar{n}$. The nonmonotonic feature of phonon distribution appears when $n_{\text{cr}}\ge 1$ such that $P(n_0)>P(n_0-1)$ as $n_0<n_{\text{cr}}$ while $P(n_0)<P(n_0-1)$ as $n_0>n_{\text{cr}}$, where $n_{\text{cr}}=N + 1 - \frac{\phi}{\chi}-\bar{n}D$. The condition $n_{\text{cr}}\ge 1$ implies a threshold of energy supply, which coincides with the one predicted by Eq.(\ref{rc}) when $D\gg 1$. Hence we can evidently conclude that the out-of-equilibrium condensate is featured by the non-thermal distribution of phonons, showing an analogy to the photon number statistics for a single mode maser rather than the atomic BEC \cite{Scully_PRL1999,Scully_book1997}.

In terms of $P(n_0)$ given by Eq.(\ref{Pn0r}), we are able to obtain the fluctuation of $n_0$ in further and the higher moments $\langle n_0^{\ell}\rangle$ will be presented elsewhere. For the energy pump appreciably above threshold, we have the condensate ratio defined as $\eta=\langle n_0\rangle/N$ and Mandel parameter $Q=\langle \Delta n_0^2\rangle/\langle n_0\rangle-1$
\begin{equation}
\begin{split}
& \eta = 1 - \frac{\phi(\phi+\chi\bar{n}D)}{\chi(D+1)(r+\phi\bar{n})}\\[0.15cm] 
& Q = \frac{(r+\phi\bar{n})\left[\phi-\chi(D+1)\right]+\phi(\bar{n}+2)(\phi+\chi\bar{n}D)}{\chi(D+1)r+\phi(\chi\bar{n}-\phi)}
\end{split}
\label{Mandel}
\end{equation}
Increasing the external energy pump $r$, the numerator of Mandel parameter $Q$ for the phonon condensate will become negative once $r \ge (\bar{n}+1)\left(\frac{2\phi}{\chi}+\bar{n}D\right)/\left(\frac{D}{\phi}-\frac{1}{\chi}\right)$ as long as $\frac{\chi}{\phi}>\frac{1}{D+1}$. Then the sub-Poissonian distribution of phonons will show up, manifesting the non-classical properties. This is supported by the numerical calculations of phonon statistics shown in SM.

Fig.\ref{phononswt300} illustrates the phonon statistics of the lowest vibrational mode $\omega_0$ of BSA protein at room temperature, with respect to various rates of energy supply. The super-Poissonian distribution is shown in Fig.\ref{phononswt300}(a) with the condensate ratio $\langle n_0\rangle/N\simeq 69\%$, when protein is pumped at the rate of $r\simeq 220$GHz above the threshold . This would be feasible since the energy pump is still not strong compared to the vibrational frequencies $\sim 0.3$THz of BSA protein. However, the threshold for sub-Poissonian statistics of phonons of BSA protein is estimated to be $r=2.98$THz, which considerably exceeds its vibrational frequencies $\sim 0.3$THz. This indicates a strong-field pump that would however cause other effects, such as anharmonicity and ionization. Hence the observation of condensate of phonons at the lowest mode would be feasible under room temperature, but the non-classical distribution will be smeared out. This is more feasible when cooling the surround medium, which is shown in Fig.1 in SM.

It should be noted that the quantum fluctuations attracted much attention in atomic BEC thanks to the experimental advance during last two decades, although the mean field treatment based on Gross-Pitaevskii equation had been well developed for over half century \cite{Pethick_book2001}. The large fluctuations especially in the vicinity of critical point often makes the mean-field theory breakdown, but is crucial for understanding the critical phenomenon, even in atomic BEC, i.e., superfluid-insulator transition and BEC-BCS crossover. The phonon-number fluctuation in the condensate would become experimentally accessible by measuring the number of spontaneously emitted photons in temporal domain. The photon statistics can be then reconstructed and has been demonstrated in recent experiments on exciton polariton \cite{Klaas_PRL2018}.

{\it Discussion and conclusion.--} To experimentally implement the Fr\"ohlich condensate, many systems were suggested, i.e., microtubules (MT) \cite{Pokorny_BEC2003,Sahu_SR2014,Hameroff_JCS1994,comments_MT}, BSA and lysozyme proteins \cite{Lundholm_SD2015,Turton_NatCommun2013,Nardecchia_2017} as the most favorite ones. But the feasibility of these samples are still in debate. First let us see BSA protein whose size is $140\mathring{\text{A}}\times 40\mathring{\text{A}}\times 40\mathring{\text{A}}$. The area which the sub-THz laser with the wavelength $\lambda\simeq 400\mu$m focus on the sample is taken as $A\simeq \lambda^2/4\simeq 40000\mu$m$^2$, in accordance to the diffraction limit. Using the parameters for Fig.\ref{n0fig}, our full quantum theory leads to the estimation of energy supply to produce the condensate ratio of 50$\%$: $P_{\omega}\simeq r\hbar\omega_0\simeq 20$pW given by $r\simeq 101$GHz. With the cross section of light scattering for resonance absorption $\sigma\sim 10^{-15}$cm$^2$, we proceed via the number of photons captured by protein molecules $N=A/\sigma\sim 4\times 10^{11}$, which yields the laser power $P\simeq P_{\omega}N\sim 8$W. This would be feasible for the recent development of high-power laser using frequency mixing \cite{Tochitsky_JOSAB2007}.

To overcome the difficulty of using very strong powers of lasers is to induce the optical excitations of some fluorochromes (i.e., Alexa488) covalently bound to each protein molecule \cite{Nardecchia_2017}. The excited fluorochromes create fluorescence, resulting in the transfer of some residue energy into the vibrational modes of protein. In practice, this method takes the advantage of avoiding the optical transition of protein and suppressing the absorption by surrounded water molecules which causes the laser-induced heating up, when pumped by the Argon laser with wavelength of 488nm. Also, such indirect pumping scheme would demand much less laser power than the one using infrared maser, due to the fact that the smaller laser spot can be achieved in optical excitation regime.

For lysozyme protein, the recent advances in both experiments and theory found the sub-THz excitation at $\omega_0=0.4\times 2\pi$THz and the damping rate $\phi\sim 1$GHz \cite{Lundholm_SD2015,Martin_JCP2017}. Due to the diffraction limit, the area which the sub-THz laser with the wavelength $\lambda\simeq 400\mu$m focus on the sample is taken as $A\simeq \lambda^2/4\simeq 40000\mu$m$^2$. Under the room temperature, the full quantum theory estimates the energy supply for the condensate ratio of 50$\%$: $P_{\omega}\simeq 4.2$pW given by $r\simeq 16$GHz. Then the number of photons captured by the sample reads $N=A/\sigma\sim 4\times 10^{11}$, yielding the laser power $P\simeq P_{\omega}N\sim 1.6$W. This would be feasible for CW THz laser \cite{Tochitsky_JOSAB2007}.


In conclusion, we have developed a full quantum statistical theory using nonequilibrium equations of motion for the Fr\"ohlich condensate. Our model goes much beyond the precedent results by providing an elegant description of the critical behavior of the phase transition towards the condensation of phonons. Besides, the model led us to the phonon-number distribution of the condensate with the energy pumping, evolving from quasi-thermal to super-Poissonian statistics. 
This yields an analogy to laser operation. Moreover, we proposed an infrared spectroscopy for experimentally probing such collective mode, evident by the remarkably sharp peak with {\it narrow} linewidth. The development of quantum statistical theory for Fr\"ohlich condensate, especially the fluctuation aspects, offers the new insights and perspective for the driven-dissipation systems that would stimulate the research on novel properties of low-frequency vibrations of the materials far-from-equilibrium.

We gratefully acknowledge the support of the grants AFOSR Award FA-9550-18-1-0141, ONR Award N00014-16-1-3054 and Robert A. Welch Foundation (Award A-1261 $\&$ Award A-1943-20180324). We also thank V. V. Yakovlev, K. Wang, Z. H. Yi, A. Sokolov and M. King for the useful discussions.


\begin{thebibliography}{}
\bibitem{comments_MT}
The longitudinal modes that are most electromagnetically active in MT's are expected to be in GHz regime and these would require much higher pumping power, making the possibility of observing Fr\"ohlich condensation in MT's somewhat remote.
\end{thebibliography}
\end{document}


\title{Quantum fluctuations in Fr\"ohlich condensate of molecular vibrations driven far from equilibrium: Supplementary Material}

\author{Zhedong Zhang}
\affiliation{Institute for Quantum Science and Engineering, Texas A$\&$M University, College Station, TX 77843, USA}

\author{Girish S. Agarwal}
\affiliation{Institute for Quantum Science and Engineering, Texas A$\&$M University, College Station, TX 77843, USA}

\author{Marlan O. Scully}
\affiliation{Institute for Quantum Science and Engineering, Texas A$\&$M University, College Station, TX 77843, USA}
\affiliation{Department of Physics, Baylor University, Waco, TX 76798, USA}
\affiliation{Department of Physics, Princeton University, Princeton, NJ 08544, USA}

\date{\today}

\begin{abstract}
This supplementary material will provide the derivations for some equations in main text and also some detail analysis of Fr\"ohlich condensate in support of the main text.
\end{abstract}

\maketitle

\section{Description of molecules immersed in solvent}
When the molecules are placed in dense medium, i.e., solvent, the full description usually includes the energy of electrons, nucleis of molecules and nucleis of the medium, as well as their interactions. In practice, however,
modeling to include so many degrees of freedoms is overwhelmly complicated, which makes the theory elusive and unaccessible. Fortunately, thank to their distinct energy and time scales, we do not have to deal with such a large amount of DOF in practice. This enables us to simplify the model which contains certain DOF of interests. Since we are focusing on the low-frequency (THz) vibrations of molecules where the electron motions will not be altered much in experiments, the effective description of the system contains only the nucleis of molecules and surrounding medium (solvent), dropping the electron energy. In this sense, the Hamiltonian of our interest reads
\begin{equation}
\begin{split}
H = H_{nul}^0(\textbf{p},\textbf{q}) + H_{sol}(\textbf{P},\textbf{Q}) + V(\textbf{q},\textbf{Q})
\end{split}
\label{Hn}
\end{equation}
where $\textbf{p},\textbf{P}$ are the respective momentums of nucleis in molecules and medium, in conjugate with their coordinates $\textbf{q},\textbf{Q}$. The 1st and 2nd terms in Eq.(\ref{Hn}) stand for the free Hamiltonians of nucleis in molecules and medium, respectively. The 3rd term is the potential energy that the nucleis in molecules feel when embedded in the medium. By treating the nuclear motion as many harmonic oscillators, the Hamiltonian is of the form
\begin{equation}
\begin{split}
& H_{nul}^0(\textbf{p},\textbf{q}) = \sum_i\left(\frac{p_i^2}{2m_i}+\frac{1}{2}m_i\omega_i^2 q_i^2\right) + \sum_k V(\textbf{q},0),\quad H_{sol}(\textbf{P},\textbf{Q}) = \sum_k\left(\frac{P_k^2}{2M_k}+\frac{1}{2}M_k\Omega_k^2 Q_k^2\right)\\[0.15cm]
& \bar{V}(\textbf{q},\textbf{Q}) = V(\textbf{q},\textbf{Q}) - V(\textbf{q},0)
\end{split}
\label{Ht}
\end{equation}
Note that the interaction $\bar{V}(\textbf{q},0)=0$. For simplicity, let us assume the additive form of interaction with respect to solvent coordinates $\textbf{Q}$, namely, $V(\textbf{q},\textbf{Q})=\sum_k V(\textbf{q},Q_k)$ so that $\bar{V}(\textbf{q},\textbf{Q})=\sum_k \bar{V}(\textbf{q},Q_k)$. Because of the small deviations of nucleis from their equilibrium positions, we proceed by keeping those terms such as $q_i Q_k$, $q_i q_j Q_k$ and $q_j Q_k^2$ through the expansion of the interaction $\bar{V}(\textbf{q},\textbf{Q})$ in terms of $q_i$ and $Q_k$. On the other hand, the two-quanta absorption/emission process governed by $q_j Q_k^2$ terms will be neglected in current work, because of its lower chance to take place than the Raman-like process given by $q_i q_j Q_k$ terms. With these concerns and Eq.(\ref{Ht}), we are led to the coupling between molecular vibrations and the surrounding dense medium (solvent) in the second quantization formalism
\begin{equation}
\begin{split}
V_{\text{ms}}(t) = \hbar\sum_i\sum_k f_{i,k}\left(a_i^{\dagger}b_k e^{i(\bar{\omega}_i-\bar{\Omega}_k)t} + \text{h.c.}\right) + \hbar\sum_{i,j}\sum_k G_{ij,k}\left(a_i^{\dagger}a_j b_k e^{i(\bar{\omega}_i-\bar{\omega}_j-\bar{\Omega}_k)t} + \text{h.c.}\right)
\end{split}
\label{Vt}
\end{equation}
which was alternatively termed as Wu-Austin Hamiltonian in the literatures \cite{Wu_JBP1981}. $f_{i,k}$ and $G_{ij,k}$ refer to the coupling constants between molecular vibrations and solvent, as determined by the geometry of potential which nuleis in molecules feel in the presence of solvent. In fact the magnitudes of these coupling constants were figured out by fitting with experiments. $a_i$ and $b_k$ are the bosonic annihilation operators for the $i$-th vibrational mode of molecules and the solvent modes, respectively. $\bar{\omega}_i$ and $\bar{\Omega}_k$ denote the respective frequencies associated with the $i$-th vibrational mode of molecules and the solvent modes.

\section{Phonon master equation}
The external source needs to provide the molecules energy input, driving the system into out-of-equilibrium regime, i.e., laser pump. In reality, the solvent surrounding the molecules always causes the dephasing between molecules, which subsequently effects in the incoherent pumping. In this sense, the coherence between different vibrational modes will not be considered when deriving the equation of motion (EOM) for phonons. The effective Hamiltonian for pumping has the following form
\begin{equation}
\begin{split}
V_{\text{p}}(t) = \hbar\sum_{s=0}^D\left(F_s^*(t)a_s e^{-i\omega_s t} + F_s(t)a_s^{\dagger} e^{i\omega_s t}\right)
\end{split}
\label{Vpump}
\end{equation}
where $F_s(t)$ denotes the effective broad-band pumping field whose spectrum is assumed to cover the low-frequency vibrations (THz) considered here $\langle F_s^*(t)F_{s'}(t')\rangle = \frac{p}{2}\delta_{ss'}\delta(t-t')$.

The solvent usually contains a large amount of nuclear degrees of freedoms, which may relax much faster than the molecules. This part of degrees of freedoms can be properly treated as Markovian bath, since we are interested in the dynamics of molecular vibrations. Using the standard method \cite{Agarwal_book2013}, the reduced master equation is given by
\begin{equation}
\begin{split}
\dot{\rho} = -\frac{1}{\hbar^2}\left(\int_0^{\infty}\text{d}t'\ \text{Tr}_b\langle[V_{\text{p}}(t),[V_{\text{p}}(t-t'),\rho_b\otimes\rho(t)]]\rangle + \int_0^{\infty}\text{d}t'\ \text{Tr}_b[V_{\text{ms}}(t),[V_{\text{ms}}(t-t'),\rho_b\otimes\rho(t)]]\right)
\end{split}
\label{cg}
\end{equation}
where the state of the full system $\rho_T(t)=\rho(t)\otimes\rho_b(0)+\rho_c$ and $\rho_c$ is of the higher order of system-bath coupling. Notice that in Eq.(\ref{cg}) $\rho(t')$ has been replaced by $\rho(t)$ under Markovian approximation. The solvent is at thermal equilibrium with finite temperature, so that $\rho_b=Z_b^{-1}\prod_k \text{exp}(-\hbar\bar{\Omega}_k b_k^{\dagger}b_k/k_B T)$. $\text{Tr}_b$ and $\langle\cdots\rangle$ denote the average over the bath and stochastic pumping field, respectively. Substituting Eq.(\ref{Vt}) into Eq.(\ref{cg}), some algebra gives rise to the equation of motion
\begin{equation}
\begin{split}
\dot{\rho} = \sum_{s=0}^D\frac{r_{\omega_s}}{2} & \left(a_s\rho a_s^{\dagger}-a_s^{\dagger}a_s\rho+a_s^{\dagger}\rho a_s-a_s a_s^{\dagger}\rho\right) + \frac{\phi}{2}\sum_{s=0}^D\left[(\bar{n}_{\omega_s}+1)\left(a_s\rho a_s^{\dagger}-a_s^{\dagger}a_s\rho\right)+\bar{n}_{\omega_s}\left(a_s^{\dagger}\rho a_s-a_s a_s^{\dagger}\rho\right)\right]\\[0.15cm]
& + \frac{\chi}{2}\sum_{s=1}^D\sum_{l=0}^{s-1}\left[(\bar{n}_{\omega_{sl}}+1)\left(a_l^{\dagger}a_s\rho a_s^{\dagger}a_l-a_s^{\dagger}a_l a_l^{\dagger}a_s\rho\right) + \bar{n}_{\omega_{sl}}\left(a_s^{\dagger}a_l\rho a_l^{\dagger}a_s-a_l^{\dagger}a_s a_s^{\dagger}a_l\rho\right)\right] + \text{h.c.}
\end{split}
\label{qme}
\end{equation}
and $\bar{n}_{\omega}=[\text{exp}(\hbar\omega/k_B T)-1]^{-1}$ is the Planck factor. $r_{\omega}=2\pi p{\cal D}(\omega)$, $\phi=2\pi f_{\omega}^2{\cal D}(\omega),\ \chi=2\pi G_{\omega}^2{\cal D}(\omega)$ are the rates of pumping, dissipation (one-phonon) and energy redistribution (two-phonon), respectively. ${\cal D}(\omega)$ is the density of states of the solvent as a bath. Due to the smooth spectral density of bath over the range of vibration frequency considered, the rates of dissipation and energy redistribution can be properly approximated to be the same for each mode. Usually the two-phonon process takes place slower than the one-phonon process, namely, $\chi<\phi$.

From Eq.(\ref{Vt}), the total number of phonons will be solely dictated by the pumping and dissipation, regardless of the energy redistribution process. To manifest this, we start from the rate equations of phonon numbers
\begin{equation}
\begin{split}
\langle \dot{n}_l\rangle & = r_l + \phi\big[\bar{n}_{\omega_l}\left(\langle n_l\rangle+1\right) - \left(\bar{n}_{\omega_l}+1\right)\langle n_l\rangle\big]\\[0.15cm]
& \ + \chi\Big\{\sum_{j>l}\big[\left(\bar{n}_{\omega_{jl}}+1\right)\langle n_j(n_l+1)\rangle - \bar{n}_{\omega_{jl}}\langle (n_j+1)n_l\rangle\big] + \sum_{j<l}\big[\bar{n}_{\omega_{lj}}\langle n_j(n_l+1)\rangle - \left(\bar{n}_{\omega_{lj}}+1\right)\langle (n_j+1)n_l\rangle\big]\Big\}
\end{split}
\label{nl}
\end{equation}
which can be obtained from Eq.(\ref{qme}). By further approximating the constant pumping as well as Planck factor $r_l\simeq r,\ \bar{n}_{\omega_l}\simeq \bar{n}_{\omega_0}\equiv\bar{n},\ \bar{n}_{\omega_{jl}}\simeq \bar{n}_{\omega_0}\equiv\bar{n}$ due to the fact of smooth spectrum density of the bath, we obtain the rate equation for the total number of phonons $N=\sum_{s=0}^D a_s^{\dagger}a_s$
\begin{equation}
\begin{split}
\langle\dot{N}\rangle & = (D+1)(r+\phi\bar{n}) - \phi\langle N\rangle + \chi\Big(\sum_{l=0}^{D-1}\sum_{j=l+1}^D\langle n_j(n_l+1)\rangle - \sum_{l=1}^D\sum_{j=0}^{l-1}\langle (n_j+1)n_l\rangle\Big)\\[0.15cm]
& = (D+1)(r+\phi\bar{n}) - \phi\langle N\rangle
\end{split}
\label{Nt}
\end{equation}
where the $\chi$-term above exactly vanishes because of the identity
\begin{equation}
\begin{split}
\sum_{l=0}^{D-1}\sum_{j=l+1}^D\langle n_j(n_l+1)\rangle = \sum_{j=1}^D\sum_{l=0}^{j-1}\langle n_j(n_l+1)\rangle\equiv \sum_{l=1}^D\sum_{j=0}^{l-1}\langle n_l(n_j+1)\rangle
\end{split}
\label{id}
\end{equation}
In what follows we can properly assume the stationary $\langle N\rangle\simeq (D+1)(r_{\omega}+\phi\bar{n}_{\omega})/\phi$ during the observation timescale $\tau\gg \phi^{-1}$.

\section{Connection to Fr\"ohlich's work}
By making the decorrelation approximation between vibrational modes, we can recast Eq.(\ref{nl}) into
\begin{equation}
\begin{split}
\langle \dot{n}_l\rangle = & r - \phi\bar{n}_{\omega_l}\big[\langle n_l\rangle e^{\hbar\omega_l/k_B T} - (\langle n_l\rangle+1)\big]\\[0.15cm]
& \ - \chi\Big\{\left(\bar{n}+1\right)\sum_{j>l}\big[\langle n_j+1\rangle\langle n_l\rangle e^{\hbar(\omega_l-\omega_j)/k_B T} - \langle n_j\rangle\langle n_l+1\rangle\big]\\[0.15cm]
& \qquad\qquad + \bar{n}\sum_{j<l}\big[\langle n_j+1\rangle\langle n_l\rangle e^{\hbar(\omega_l-\omega_j)/k_B T} - \langle n_j\rangle\langle n_l+1\rangle\big]\Big\}
\end{split}
\label{nlf}
\end{equation}
which recovers Fr\"ohlich's results \cite{Frohlich_IJQC1968}. Based on Eq.(\ref{nl}), the rate equation for $n_0$ is given by
\begin{equation}
\begin{split}
\langle \dot{n}_0\rangle = r + \phi\big[\bar{n}(\langle n_0\rangle+1) - (\bar{n}+1)\langle n_0\rangle\big] + \chi\sum_{s=1}^D\big[(\bar{n}_{\omega_{s0}}+1)\langle n_s\rangle(\langle n_0\rangle+1) - \bar{n}_{\omega_{s0}}(\langle n_s\rangle+1)\langle n_0\rangle\big]
\end{split}
\label{n0}
\end{equation}
where in the $\chi$-part in Eq.(\ref{n0}) the first term describes the process such that the mode $\omega_0$ is pumped by the mode $\omega_s$, together with the dissipation by the bath mode $\omega_s-\omega_0$. The reverse process is described by the second term. These are in correspondence to the Feynman diagrams shown in Fig.1(c) in the main text. At the steady state, the phonon number at mode $\omega_0$ can be formally solved as
\begin{equation}
\langle n_0\rangle = \frac{r+\phi\bar{n}+\chi(\bar{n}+1)\langle N_e\rangle}{\phi-\chi(\langle N_e\rangle-D\bar{n})}
\label{n0sol}
\end{equation}
by introducing the phonon number at higher-frequency modes $\langle N_e\rangle=\sum_{l=1}^D\langle a_l^{\dagger}a_l\rangle$ and the positivity of $\langle n_0\rangle$ results in the upper bound of $\langle N_e\rangle$
\begin{equation}
\langle N_e\rangle < \frac{\phi}{\chi} + D\bar{n}
\label{Nub}
\end{equation}
Since the total number of phonons $\langle N\rangle$ is solely determined by pumping and dissipation (one-phonon processes) as shown in Eq.(\ref{Nt}), the condensation of phonons at mode $\omega_0$ will therefore take place once the external energy supply $r_{\omega}$ exceeds a threshold. 
To quantify this, we recast Eq.(\ref{n0}) by inserting $\langle N_e\rangle=\langle N\rangle-\langle n_0\rangle$
\begin{equation}
\begin{split}
& \langle \dot{n}_0\rangle = a(r-r_{\text{c}})\langle n_0\rangle - \chi\langle n_0\rangle^2 + b(r+\phi\bar{n})\\[0.15cm]
& r_{\text{c}} = \frac{\phi}{D+1}\left(1+\frac{\phi}{\chi}\right),\quad a = \frac{(D+1)\chi}{\phi},\quad b = 1 + (D+1)(\bar{n}+1)\frac{\chi}{\phi}
\end{split}
\label{n0ls}
\end{equation}
which is a reminiscence of laser equation, in that $a(r-r_{\text{c}})$ serves as gain that becomes positive when the pumping exceeds the threshold $r_{\text{c}}$ and $\chi$ gives the saturation (nonlinear). The constant term $b(r+\phi\bar{n})$ results in the residue number of phonons at mode $\omega_0$ below the threshold, which is analogous with the equilibrium Bose-Einstein condensate. Notice that in deriving Eq.(\ref{n0ls}) the total number of phonons $\langle N\rangle$ has been replaced by its stationary value as given by Eq.(\ref{Nt}). This takes the rational when considering the physics within timescale $\tau\gg\phi^{-1}$.


\subsection{POPULATION DYNAMICS OF THE LOWEST MODE}
Obviously, the two-phonon process governed by the 2nd term in Eq.(\ref{Vt}) commutes with the total phonon number operator $N$. This, in other words, implies that the total phonon number always keeps unchanged under two-phonon process. In this sense, the two-phonon process would results in the correlation between the fluctuations of $n_0$ and $n_s (s\ge 1)$. To manifest this, we essentially consider the physics under the timescale $t\gg\phi^{-1}$ and fast pumping so that the total phonon number properly reaches its stationary value
\begin{equation}
\begin{split}
\langle N\rangle = (D+1)\left(\frac{r}{\phi} + \bar{n}\right)
\end{split}
\label{Ntss}
\end{equation}
given by Eq.(\ref{Nt}), which shows that the total number of phonons is solely dictated by one-phonon processes, regardless of two-phonon process. Then the identity
\begin{equation}
\begin{split}
\langle n_0\rangle + \sum_{s=1}^D\langle a_s^{\dagger}a_s\rangle = \langle N\rangle
\end{split}
\label{idn}
\end{equation}
indicates a constraint between the phonon numbers at mode $\omega_0$ and higher-frequency modes $\omega_s(s\ge 1)$. The correlation between the phonon numbers at mode $\omega_0$ and modes $\omega_s(s\ge 1)$ then becomes important, in that the two-phonon process may annihilate(create) one phonon at mode $\omega_0$ and subsequently create(annihilate) one phonon at modes $\omega_s(s\ge 1)$, keeping $\langle N\rangle$ invariant. To capture this feature, it would be better to assume decorrelation between $n_0$ and $N$ such that $\langle n_0 N\rangle\simeq \langle n_0\rangle\langle N\rangle$, rather than the decorrelation approximation made for all the vibrational modes $\langle n_i n_j\rangle\simeq \langle n_i\rangle\langle n_j\rangle,\ (i,j=0,1,2,\cdots,D)$. To this end, we define the density matrix for the mode $\omega_0$
\begin{equation}
\begin{split}
\rho_{n_0,m_0} = \sum_{\{n_l\}}\langle n_0,\{n_l\}|\rho|m_0,\{n_l\}\rangle,\quad \{n_l\} = \{n_1,n_2,\cdots,n_D\}
\end{split}
\label{rho0}
\end{equation}
and notes the fact
\begin{equation}
\begin{split}
& \sum_{\{n_l\}}\langle n_0,\{n_l\}|\left(2a_0\rho a_0^{\dagger}-a_0^{\dagger}a_0\rho-\rho a_0^{\dagger}a_0\right)|n_0,\{n_l\}\rangle\\[0.15cm]
& = 2\sum_{\{n_l\}}\Big((n_0+1)\langle n_0+1;\{n_l\}|\rho|n_0+1;\{n_l\}\rangle - n_0\langle n_0;\{n_l\}|\rho|n_0;\{n_l\}\rangle\Big)\\[0.15cm]
& = 2\Big((n_0+1)\rho_{n_0+1,n_0+1}-n_0\rho_{n_0,n_0}\Big)\\[0.15cm]
& \sum_{\{n_l\}}\langle n_0,\{n_l\}|\left(2a_0^{\dagger}\rho a_0-a_0 a_0^{\dagger}\rho-\rho a_0 a_0^{\dagger}\right)|n_0,\{n_l\}\rangle\\[0.15cm]
& = 2\sum_{\{n_l\}}\Big(n_0\langle n_0-1;\{n_l\}|\rho|n_0-1;\{n_l\}\rangle - (n_0+1)\langle n_0;\{n_l\}|\rho|n_0;\{n_l\}\rangle\Big)\\[0.15cm]
& = 2\Big(n_0\rho_{n_0-1,n_0-1} - (n_0+1)\rho_{n_0,n_0}\Big)\\[0.15cm]
& \sum_{\{n_l\}}\langle n_0,\{n_l\}|\left(2a_s\rho a_s^{\dagger}-a_s^{\dagger}a_s\rho-\rho a_s^{\dagger}a_s\right)|n_0,\{n_l\}\rangle\\[0.15cm]
& = 2\sum_{\{n_l\}}\Big((n_s+1)\langle n_0;\{n_1,\cdots,n_s+1,\cdots,n_D\}|\rho|n_0;\{n_1,\cdots,n_s+1,\cdots,n_D\}\rangle - n_s\langle n_0;\{n_l\}|\rho|n_0;\{n_l\}\rangle\Big)\\[0.15cm]
& = 0\\[0.15cm]
& \sum_{\{n_l\}}\langle n_0,\{n_l\}|\left(2a_s^{\dagger}\rho a_s-a_s a_s^{\dagger}\rho-\rho a_s a_s^{\dagger}\right)|n_0,\{n_l\}\rangle\\[0.15cm]
& = 2\sum_{\{n_l\}}\Big(n_s\langle n_0;\{n_1,\cdots,n_s-1,\cdots,n_D\}|\rho|n_0;\{n_1,\cdots,n_s-1,\cdots,n_D\}\rangle - (n_s+1)\langle n_0;\{n_l\}|\rho|n_0;\{n_l\}\rangle\Big)\\[0.15cm]
& = 0
\end{split}
\label{dd}
\end{equation}
which gives
\begin{equation}
\begin{split}
\dot{\rho}_{n_0,n_0} = & \left(r+\phi(\bar{n}+1)\right)\Big((n_0+1)\rho_{n_0+1,n_0+1} - n_0\rho_{n_0,n_0}\Big) + \left(r+\phi\bar{n}\right)\Big(n_0\rho_{n_0-1,n_0-1} - (n_0+1)\rho_{n_0,n_0}\Big)\\[0.15cm]
& + \chi\sum_{s=1}^D\bigg[(\bar{n}_{\omega_{s0}}+1)\sum_{\{n_l\}}\Big(n_0 n_s\langle n_0-1;\{n_l\}|\rho|n_0-1;\{n_l\}\rangle - (n_0+1)n_s\langle n_0;\{n_l\}|\rho|n_0;\{n_l\}\rangle\Big)\\[0.15cm]
& \qquad\ + \bar{n}_{\omega_{s0}}\sum_{\{n_l\}}\Big((n_0+1)(n_s+1)\langle n_0+1;\{n_l\}|\rho|n_0+1;\{n_l\}\rangle - n_0(n_s+1)\langle n_0;\{n_l\}|\rho|n_0;\{n_l\}\rangle\Big)\bigg]
\end{split}
\label{rhol}
\end{equation}
Using the identity
\begin{equation}
\begin{split}
\langle n_0;\{n_l\}|\rho|n_0;\{n_l\}\rangle = P(\{n_l\}|n_0)\rho_{n_0,n_0}
\end{split}
\label{jp}
\end{equation}
from the probability theory, where $P(\{n_l\}|n_0)$ stands for the probability of phonon configuration $\{n_l\}$ given that $n_0$ phonons at mode $\omega_0$, one has
\begin{equation}
\begin{split}
& \sum_{\{n_l\}}n_s\langle m_0;\{n_l\}|\rho|m_0;\{n_l\}\rangle = \sum_{\{n_l\}}n_s P(\{n_l\}|m_0)\rho_{m_0,m_0} = \langle n_s\rangle_{m_0}\rho_{m_0,m_0}\\[0.15cm]
& \sum_{\{n_l\}}(n_s+1)\langle m_0;\{n_l\}|\rho|m_0;\{n_l\}\rangle = \sum_{\{n_l\}}(n_s+1)P(\{n_l\}|m_0)\rho_{m_0,m_0} = \langle n_s+1\rangle_{m_0}\rho_{m_0,m_0}
\end{split}
\label{jp1}
\end{equation}
where $m_0=n_0,n_0\pm 1$. Inserting Eq.(\ref{jp1}) into Eq.(\ref{rhol}), some algebra leads to the dynamical equation which the population at mode $\omega_0$ obeys
\begin{equation}
\begin{split}
\dot{\rho}_{n_0,n_0} = & - \big(r+\phi\bar{n}+\chi{\cal N}_{n_0}\big)(n_0+1)\rho_{n_0,n_0} + \big(r+\phi\bar{n}+\chi{\cal N}_{n_0-1}\big)n_0\rho_{n_0-1,n_0-1}\\[0.15cm]
& - \big(r+\phi(\bar{n}+1)+\chi{\cal M}_{n_0}\big)n_0\rho_{n_0,n_0} + \big(r+\phi(\bar{n}+1)+\chi{\cal M}_{n_0+1}\big)(n_0+1)\rho_{n_0+1,n_0+1}
\end{split}
\label{re}
\end{equation}
with
\begin{equation}
\begin{split}
{\cal N}_{n_0} = \sum_{s=1}^D(\bar{n}_{\omega_{s0}}+1)\langle n_s\rangle_{n_0},\quad {\cal M}_{n_0} = \sum_{s=1}^D\bar{n}_{\omega_{s0}}\langle n_s+1\rangle_{n_0}
\end{split}
\label{RT}
\end{equation}
${\cal N}_{n_0}$ and ${\cal M}_{n_0}$ can be evaluated in varying degrees of rigor. Here we choose the one such that $\bar{n}_{\omega_{s0}}\simeq \bar{n}_{\omega_0}\equiv\bar{n}$ which leads to ${\cal N}_{n_0}\simeq(\bar{n}+1)(\langle N\rangle-n_0),\ {\cal M}_{n_0}\simeq\bar{n}(\langle N\rangle-n_0+D)$ and then Eq.(2) in the main text can be recovered. The physical meaning of Eq.(\ref{re}) can be interpreted in terms of probability flows as shown in Fig.2 in the main text. Moreover, we obtain the rate equation for phonon number at mode $\omega_0$
\begin{equation}
\begin{split}
\langle \dot{n}_0\rangle = a(r-r_{\text{c}})\langle n_0\rangle - \chi\langle n_0^2\rangle + b(r+\phi\bar{n})
\end{split}
\label{n0lsf}
\end{equation}
with the same definition of $a,b$ and $r_{\text{c}}$ as in Eq.(\ref{n0ls}).

\subsection{OFF-DIAGONAL ELEMENTS OF THE LOWEST MODE}
The coherence $\rho_{n_0,m_0}\ (m_0\neq n_0)$ requires different numbers of phonons between the two states, according to Eq.(\ref{rho0}). This is to say that one has to introduce the fluctuation of phonon number $\langle N\rangle$, which takes the rational in that the dynamical process is under consideration when discussing the coherence lifetime. The dynamical equation of coherence then reads
\begin{equation}
\begin{split}
& \dot{\rho}_{n_0,n_0+1} = i\omega_0\rho_{n_0,n_0+1} + \big(r+\phi\bar{n}\big)\left[\sqrt{n_0(n_0+1)}\ \rho_{n_0-1,n_0} - \left(n_0+\frac{3}{2}\right)\rho_{n_0,n_0+1}\right]\\[0.15cm]
& \quad\ + \big(r+\phi(\bar{n}+1)\big)\left[\sqrt{(n_0+1)(n_0+2)}\ \rho_{n_0+1,n_0+2} - \left(n_0+\frac{1}{2}\right)\rho_{n_0,n_0+1}\right]\\[0.15cm]
& \quad\ + \chi\sum_{s=1}^D\bigg\{(\bar{n}_{\omega_{s0}}+1)\sum_{\{n_l\}}n_s\left[\sqrt{n_0(n_0+1)}\langle n_0-1;\{n_l\}|\rho|n_0;\{n_l\}\rangle - \left(n_0+\frac{3}{2}\right)\langle n_0;\{n_l\}|\rho|n_0+1;\{n_l\}\rangle\right]\\[0.15cm]
& \qquad + \bar{n}_{\omega_{s0}}\sum_{\{n_l\}}(n_s+1)\left[\sqrt{(n_0+1)(n_0+2)}\langle n_0+1;\{n_l\}|\rho|n_0+2;\{n_l\}\rangle - \left(n_0+\frac{1}{2}\right)\langle n_0;\{n_l\}|\rho|n_0+1;\{n_l\}\rangle\right]\bigg\}
\end{split}
\label{rhooff}
\end{equation}
by inserting Eq.(\ref{rho0}) into Eq.(\ref{qme}) and $\langle N\rangle=\sum_{s=0}^D n_s$. $\rho_{n_0,n_0\pm 1}$ is the one that mainly contributes to the condensate in response to the external field, because the dipole-induced single photon absorption/emission dominates. By invoking $\bar{n}_{\omega_{s0}}\simeq \bar{n}_{\omega_0}\equiv\bar{n}$ and making use of the relations
\begin{equation}
\begin{split}
& \sum_{s=1}^D\sum_{\{n_l\}}n_s\langle n_0-1;\{n_l\}|\rho|n_0;\{n_l\}\rangle = (\langle N\rangle-n_0+1)\rho_{n_0-1,n_0}\\[0.15cm]
& \sum_{s=1}^D\sum_{\{n_l\}}n_s\langle n_0;\{n_l\}|\rho|n_0+1;\{n_l\}\rangle = (\langle N\rangle-n_0)\rho_{n_0,n_0+1}\\[0.15cm]
& \sum_{s=1}^D\sum_{\{n_l\}}(n_s+1)\langle n_0+1;\{n_l\}|\rho|n_0+2;\{n_l\}\rangle = (\langle N\rangle-n_0-1+D)\rho_{n_0+1,n_0+2}\\[0.15cm]
& \sum_{s=1}^D\sum_{\{n_l\}}(n_s+1)\langle n_0;\{n_l\}|\rho|n_0+1;\{n_l\}\rangle = (\langle N\rangle-n_0+D)\rho_{n_0,n_0+1}
\end{split}
\label{id1}
\end{equation}
and Eq.(\ref{rhooff}) the equation which $\rho_{n_0,n_0+1}$ obeys takes the form of
\begin{equation}
\begin{split}
\dot{\rho}_{n_0,n_0+1} = (i\omega_0-\gamma_{n_0})\rho_{n_0,n_0+1} + c_{n_0-1}\rho_{n_0-1,n_0} - (c_{n_0}+d_{n_0})\rho_{n_0,n_0+1} + d_{n_0+1}\rho_{n_0+1,n_0+2}
\end{split}
\label{rhooffn}
\end{equation}
where
\begin{equation}
\begin{split}
& \gamma_{n_0} = \frac{r+\phi(\bar{n}+1)+\chi\bar{n}(\langle N\rangle-n_0+D)}{4\left(\sqrt{n_0(n_0+1)}+n_0\right)+2} + \frac{r+\phi\bar{n}+\chi(\bar{n}+1)(\langle N\rangle-n_0)}{4\left(\sqrt{(n_0+1)(n_0+2)}+n_0\right)+6}\\[0.2cm]
& c_{n_0} = \sqrt{(n_0+1)(n_0+2)}\left[r+\phi\bar{n}+\chi(\bar{n}+1)(\langle N\rangle-n_0)\right]\\[0.2cm]
& d_{n_0} = \sqrt{n_0(n_0+1)}\left[r+\phi(\bar{n}+1)+\chi\bar{n}(\langle N\rangle-n_0+D)\right]
\end{split}
\label{gamma0}
\end{equation}
We are facing a three-term differential recurrence of complicated but time-independent coefficients, since each term $\rho_{n_0,n_0+1}$ couples only its nearest neighbors, i.e., $\rho_{n_0-1,n_0}$ and $\rho_{n_0+1,n_0+2}$. To solve Eq.(\ref{rhooffn}) we impose the detailed balance into the initial condition $c_{n_0}\rho_{n_0,n_0+1}(0)=d_{n_0+1}\rho_{n_0+1,n_0+2}(0)$ (analogously $c_{n_0-1}\rho_{n_0-1,n_0}(0)=d_{n_0}\rho_{n_0,n_0+1}(0)$) which suggests the ansatz for the solution \cite{Scully_book1997}
\begin{equation}
\begin{split}
& \rho_{n_0,n_0+1}(t) = e^{[i\omega_0 t-D_{n_0}(t)]}\rho_{n_0,n_0+1}(0) = e^{[i\omega_0 t-D_{n_0}(t)]}\rho_{01}(0)\prod_{m=1}^{n_0}\frac{c_{m-1}}{d_m}\\[0.15cm]
& \dot{D}_{n_0} = \gamma_{n_0} + d_{n_0}\left(1-e^{-(D_{n_0-1}-D_{n_0})}\right) + c_{n_0}\left(1-e^{-(D_{n_0+1}-D_{n_0})}\right)
\end{split}
\label{Dp}
\end{equation}
One can further expand the exponential by assuming $|D_{n_0-1}-D_{n_0}|\ll 1$ and thus $D_{n_0}(t)\simeq \gamma_{n_0}t$. This holds when $\gamma_{n_0}$ is a slowly varying function of $t$, because of
\begin{equation}
\begin{split}
|D_{n_0-1}-D_{n_0}|\simeq |\gamma_{n_0-1}-\gamma_{n_0}|t\simeq \left|\frac{\partial\gamma_{n_0}}{\partial t}\right|t
\end{split}
\label{dr}
\end{equation}
and we can thereby safely replace $n_0$ in $\gamma_{n_0}$ by $\langle n_0\rangle$. Hence the vibrational coherence of the lowest mode is approximated to
\begin{equation}
\begin{split}
\rho_{n_0,n_0+1}(t) \simeq e^{(i\omega_0-\gamma_0) t}\rho_{01}(0)\prod_{m=1}^{n_0}\frac{c_{m-1}}{d_m}
\end{split}
\label{cohv}
\end{equation}
where the line-width is given by
\begin{equation}
\begin{split}
\gamma_0 & = \frac{r+\phi(\bar{n}+1)+\chi\bar{n}(\langle N\rangle-\langle n_0\rangle+D)}{4\left(\sqrt{\langle n_0\rangle(\langle n_0\rangle+1)}+\langle n_0\rangle\right)+2} + \frac{r+\phi\bar{n}+\chi(\bar{n}+1)(\langle N\rangle-\langle n_0\rangle)}{4\left(\sqrt{(\langle n_0\rangle+1)(\langle n_0\rangle+2)}+\langle n_0\rangle\right)+6}\\[0.15cm]
& \simeq \frac{r+\phi\left(\bar{n}+\frac{1}{2}\right)}{4\langle n_0\rangle}\quad \text{when}\ r_0\gg r_{\text{c}}
\end{split}
\label{decay}
\end{equation}
which gives Eq.(2) in the main text.

\section{Fluorescence spectra and line-width}
Since the absorption and emission will simultaneously take place when the damping oscillator is driven by external force, we should better collect the fluorescence signal of molecules off the forward direction to eliminate the absorption process. From the standard definition of fluorescence signal $\frac{\text{d}}{\text{d}t}\langle a_{\textbf{k}\sigma}^{\dagger}a_{\textbf{k}\sigma}\rangle$, one has
\begin{equation}
\begin{split}
S_{\text{FL}}(\omega_{\textbf{k}\sigma}) = \frac{2C{\cal E}_{\textbf{k}}}{\hbar^2}\text{Im}\left[\int_0^{\infty}\text{d}t\ e^{i\omega_{\textbf{k}\sigma} t}\text{Tr}\left(\mu^{(+)}(t)\mu^{(-)}(0)\rho_0\right)\right] = \frac{2C{\cal E}_{\textbf{k}}}{\hbar^2}\sum_{j=0}^D\frac{|\boldsymbol{\mu}_j\cdot\hat{e}_{\textbf{k}\sigma}|^2\gamma_j\langle n_j\rangle}{(\omega_{\textbf{k}\sigma}-\omega_j)^2+\gamma_j^2}
\end{split}
\label{SFL}
\end{equation}
where ${\cal E}_k\equiv\sqrt{\hbar\omega_{\textbf{k}\sigma}/2\epsilon_0 V}$ and $V$ is the bulk volumn. $C$ stands for the concentration of molecules.

\section{Phonon statistics of Fr\"ohlich condensate}
\begin{figure}
 \captionsetup{justification=raggedright,singlelinecheck=false}
 \centering
   \includegraphics[scale=0.2]{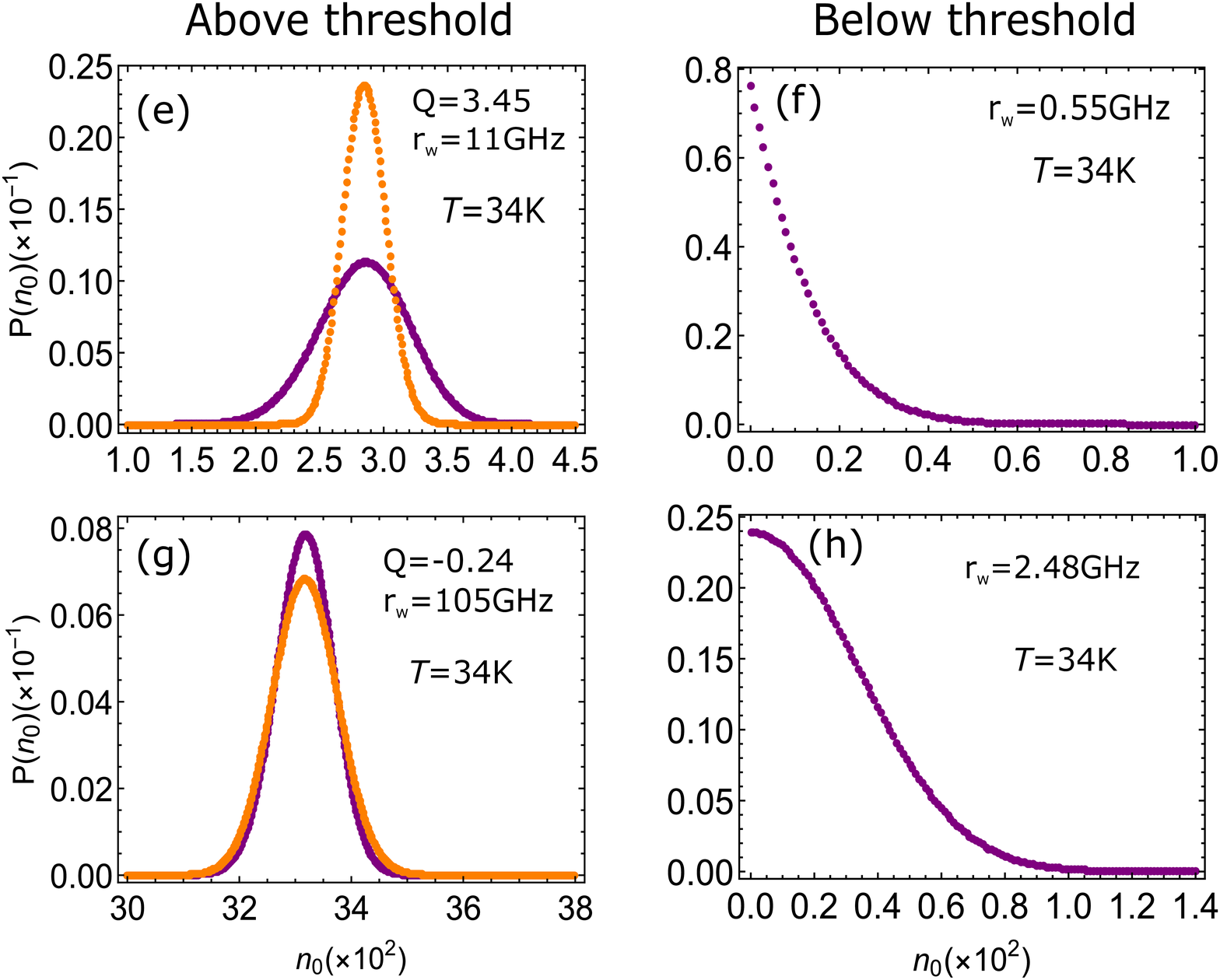}
\caption{Phonon distribution of the lowest vibrational mode where the solvent temperature is $T=34$K corresponding to $\bar{n}=1.5$. Energy pump (a) $r=11$GHz, (b) $r=0.55$GHz, (c) $r=105$GHz and (d) $r=2.48$GHz; Orange dots correspond to Poissonian distribution. Parameters are taken from Ref.\cite{Nardecchia_2017}: $\omega_0=0.314\times 2\pi$THz, $\phi=6$GHz, $\chi=0.07$GHz and $D=200$.}
\label{phononswt34}
\end{figure}

The equation of motion for the phonon distribution function $P(n_0)$ at steady state reduces to 
\begin{equation}
\begin{split}
& -(n_0+1)\left(\mathscr{X}-\alpha(n_0+1)\right)P(n_0) + n_0\left(\mathscr{X}-\alpha n_0\right)P(n_0-1)\\[0.15cm]
& \qquad - n_0\left(\mathscr{Y}-\beta n_0\right)P(n_0) + (n_0+1)\left(\mathscr{Y}-\beta(n_0+1)\right)P(n_0+1) = 0
\end{split}
\label{Pn0ss}
\end{equation}
with
\begin{equation}
\begin{split}
& \mathscr{X} = r+\phi\bar{n} + \chi(\bar{n}+1)(N+1),\quad \mathscr{Y} = r + \phi(\bar{n}+1) + \chi\bar{n}(N+D)\\[0.15cm]
& \alpha = \chi(\bar{n}+1),\quad \beta = \chi\bar{n}
\end{split}
\label{AB}
\end{equation}
The detailed balance condition makes the second-order difference equation Eq.(\ref{Pn0ss}) reduce to the equivalent system of two first-order difference equations
\begin{equation}
\begin{split}
& n_0\left(\mathscr{X}-\alpha n_0\right)P(n_0-1) - n_0\left(\mathscr{Y}-\beta n_0\right)P(n_0) =0\\[0.15cm]
& (n_0+1)\left(\mathscr{X}-\alpha(n_0+1)\right)P(n_0) - (n_0+1)\left(\mathscr{Y}-\beta(n_0+1)\right)P(n_0+1) = 0
\end{split}
\label{1or}
\end{equation}
whose solution is
\begin{equation}
\begin{split}
& P(n_0) = P(0)\left(\frac{\alpha}{\beta}\right)^{n_0}\frac{\Gamma\left(\frac{\mathscr{X}}{\alpha}\right)\Gamma\left(\frac{\mathscr{Y}}{\beta}-n_0\right)}{\Gamma\left(\frac{\mathscr{X}}{\alpha}-n_0\right)\Gamma\left(\frac{\mathscr{Y}}{\beta}\right)}\\[0.15cm]
& P(0) = \bigg[{}_2F_1\left(1,1-\frac{\mathscr{X}}{\alpha};1-\frac{\mathscr{Y}}{\beta};\frac{\alpha}{\beta}\right) - \left(\frac{\alpha}{\beta}\right)^{N+1}\frac{\Gamma\left(\frac{\mathscr{X}}{\alpha}\right)\Gamma\left(\frac{\mathscr{Y}}{\beta}-1-N\right)}{\Gamma\left(\frac{\mathscr{X}}{\alpha}-1-N\right)\Gamma\left(\frac{\mathscr{Y}}{\beta}\right)}\\[0.15cm]
& \qquad\qquad\qquad\qquad\qquad\qquad\qquad \times {}_2F_1\left(1,2+N-\frac{\mathscr{X}}{\alpha};2+N-\frac{\mathscr{Y}}{\beta};\frac{\alpha}{\beta}\right)\bigg]^{-1}
\end{split}
\label{Pn0r}
\end{equation}
where ${}_2F_1(a,b;c;z)$ denotes the hypergeometric function of order $(2,1)$. With some manipulations we obtain the mean and the 2nd moment of phonon number
\begin{equation}
\begin{split}
& \langle n_0\rangle = \frac{\alpha}{\alpha-\beta}\left[\frac{\mathscr{X}}{\alpha}-1-\frac{\mathscr{Y}}{\alpha}\left(1-P(0)\right)-P(0)\left(\frac{\alpha}{\beta}\right)^{N}\frac{\Gamma\left(\frac{\mathscr{X}}{\alpha}\right)\Gamma\left(\frac{\mathscr{Y}}{\beta}-N\right)}{\Gamma\left(\frac{\mathscr{X}}{\alpha}-1-N\right)\Gamma\left(\frac{\mathscr{Y}}{\beta}\right)}\right]\\[0.15cm]
& \langle n_0^2\rangle = \frac{\alpha}{\alpha-\beta}\left[\left(\frac{\mathscr{X}-\mathscr{Y}-\beta}{\alpha}-1\right)\langle n_0\rangle + \frac{\mathscr{Y}}{\alpha}\left(1-P(0)\right) - P(0) N\left(\frac{\alpha}{\beta}\right)^{N}\frac{\Gamma\left(\frac{\mathscr{X}}{\alpha}\right)\Gamma\left(\frac{\mathscr{Y}}{\beta}-N\right)}{\Gamma\left(\frac{\mathscr{X}}{\alpha}-1-N\right)\Gamma\left(\frac{\mathscr{Y}}{\beta}\right)}\right]
\end{split}
\label{n0v}
\end{equation}
When the energy pump appreciably exceeds threshold, the terms in Eq.(\ref{n0v}) associated with $P(0)$ are insignificant, because of $P(0)\ll 1$. Thus the mean phonon number and Mandel parameter $Q=(\langle n_0^2\rangle-\langle n_0\rangle^2)/\langle n_0\rangle-1$ read
\begin{equation}
\begin{split}
\langle n_0\rangle \simeq \frac{\mathscr{X}-\mathscr{Y}-\alpha}{\alpha-\beta},\quad Q \simeq \frac{\mathscr{Y}}{\mathscr{X}-\mathscr{Y}-\alpha} - \frac{\alpha}{\alpha-\beta}
\end{split}
\label{n0abt}
\end{equation}
which subsequently gives Eq.(9) in the main text. In support of the results in main text, the phonon distribution is shown in Fig.\ref{phononswt34} for a variety of parameters.

When cooling the environment to $T=34$K, the quantum nature of this out-of-equilibrium condensate will become feasible to be observed, since the threshold given by negative $Q$ for sub-Poissonian distribution reduces to $r=61.3$GHz which is much lower than the vibrational frequencies $\sim 0.3$THz of BSA protein. Fig.\ref{phononswt34}(b) shows the thermal-like distribution of phonons below the threshold, which gives the classical behavior of the fields. When the BSA protein is pumped at the rate of $r=11$GHz above the threshold, the distribution shows the super-Poissonian feature with the condensate ratio $\langle n_0\rangle/N_T\simeq 42\%$, as displayed in Fig.\ref{phononswt34}(a). The sub-Poissonian statistics with condensate ratio $\langle n_0\rangle/N_T\simeq 90\%$ is further manifested when the energy pump $r=105$GHz, as elucidated in Fig.\ref{phononswt34}(c). Such quantum effect seems novel and surprising since the energy redistribution as a typical nonlinear process could results in the {\it non-classical} feature of the phonon condensate when the molecules are pumped and dissipated {\it incoherently}. The similar quantum effect induced by far-from-equilibrium was also predicted before \cite{Zhang_SR2016,Zhang_JPCB2015,Zhang_JCP2014}. 
Hence a hidden quantum nature of Fr$\ddot{\text{o}}$hlich condensate is indicated and might cause the long-range coherence of the condensate, in an analogy to the so-called off-diagonal long-range order (ODLRO) proposed in superconductivity and superfluidity \cite{Yang_RMP1962}.